# Quantum technology 2.0 – topics and contributing countries from 1980 to 2018

Thomas Scheidsteger [1,] Robin Haunschild[1], Lutz Bornmann[2] and Christoph Ettl[2]

[1] *{t.scheidsteger,r.haunschild}@fkf.mpg.de*
Max Planck Institute for Solid State Research, Heisenbergstr. 1, 70659 Stuttgart (Germany)

[2] *{lutz.bornmann,christoph.ettl}@gv.mpg.det*
Administrative Headquarters of the Max Planck Society, Hofgartenstr.8, 80539 Munich (Germany)

## Abstract

The second quantum technological revolution started around 1980 with the control of single quantum particles and their interaction on an individual basis. These experimental achievements enabled physicists and engineers to utilize long-known quantum features - especially superposition and entanglement of single quantum states - for a whole range of practical applications. We use a publication set of 54,598 papers from the Web of Science published between 1980 and 2018 to investigate the time development of four main subfields of quantum technology in terms of numbers and shares of publication as well as the occurrence of topics and their relation to the 25 top contributing countries. Three successive time periods are distinguished in the analyses by their short doubling times in relation to the whole Web of Science. The periods can be characterized by the publication of pioneering works, the exploration of research topics, and the maturing of quantum technology, respectively. Compared to the US, China has a far over proportional contribution to the worldwide publication output, but not in the segment of highly-cited papers.

## Introduction

In the beginning of the 20th century Planck's quantum hypothesis to derive the correct black body radiation (Planck, 1901) and Einstein's explanation of the photoelectric effect (Einstein, 1905) led to a full-grown quantum theory in the mathematical formulations of the matrix mechanics of Heisenberg, Born, and Jordan (Born, Heisenberg, & Jordan, 1926) as well as of Schrödinger's wave mechanics. Quantum theory turned out to be highly consistent with experiment. The theory formed the basis for the development of solid state physics and for a first quantum technological revolution. This development led to applications such as lasers, transistors, nuclear power plants, solar cells, and superconducting magnets in NMR devices or particle accelerators. These applications have in common the exploitation of quantum behavior of great ensembles of particles.

In the late 1970s and early 1980s, scientists learned to prepare and to control systems of single quantum particles such as atoms, electrons, and photons, and to let the particles interact on an individual basis. This ability sparked a second quantum revolution, where physicists and engineers worked together to utilize the long-known quantum features - especially superposition and entanglement of single quantum states - for a whole range of practical "next generation" applications. These applications may be summarized as "quantum engineering" or "quantum technology 2.0" (QT 2.0).

The present study provides a bibliometric analysis of QT 2.0 methodologically following previous studies which have dealt with research fields such as climate change in general (Haunschild, Bornmann, & Marx, 2016), specific aspects thereof (Marx, Haunschild, & Bornmann, 2017a; Marx, Haunschild, & Bornmann, 2017b, 2018), and density functional theory (Haunschild, Barth, & Marx, 2016). This study analyzes QT 2.0 over the time period 1980-2018 with a focus on the following four topical fields:

(i) *Quantum information science* (Q INFO): The proposition by Einstein, Podolsky, and Rosen (1935) that quantum systems can exhibit non-local, entangled correlations unknown in the classical world could be experimentally proven after 45 years (Aspect, Grangier, & Roger, 1982; Clauser & Shimony, 1978). Since then it is the basis of quantum information processing, e.g., by the use of non-local photon correlations to make an unbreakable quantum cryptographic key distribution over a long optical fibre (Tapster, Rarity, & Owens, 1994). The concept of a quantum bit or qubit (the quantum mechanical generalization of a classical bit) can be physically realized as a two-state device, exploiting the coherent superposition of both states. The engineering challenge is the layout of hardware systems that are able to handle many qubits in quantum gates, registers and circuits. In particular, the qubits must be stored and kept stable enough to perform several computation cycles in order to realize a quantum computer.

*(ii) Quantum metrology and sensing* (Q METR): New measurement techniques provide higher precision than the same measurement performed in a classical framework. One example is a new generation of quantum logic clocks achieving a previously unknown accuracy by exploiting the sensitivity of quantum entanglement against disturbances. Other examples are atom interferometry-based gravimeters (Snadden, McGuirk, Bouyer, Haritos, & Kasevich, 1998) and magnetic field sensors based on quantum defects in diamonds (Barry et al., 2016). Control of quantum systems is achieved via manipulating quantum interferences of the wave functions in coherent laser beams. The control is guided by the so-called quantum optimal control theory (Brif, Chakrabarti, & Rabitz, 2010).

*(iii) Quantum communication and cryptography* (Q COMM): The field started by the publication of the BB84 protocol for quantum key exchange by Bennett and Brassard (1984). BB84's main component is the quantum key distribution via entangled qubits which would render under cover eavesdropping impossible. Quantum networks consist of quantum processors which exchange qubits over quantum communication channels. Secure communication in quantum networks is essential for the long-range transmission of quantum information usually by quantum teleportation.

*(iv) Quantum computing* (Q COMP): This field promises a quantum leap in computational power, since previous speed-ups as described by Moore's law (Moore, 1995) appear to come to an end on the basis of semiconductor technology (Bilal, Ahmed, & Kakkar, 2018). The original idea of quantum computing had been expressed by Feynman (1982): quantum systems as, e.g., molecules should be simulated by letting a model quantum system evolve and calculate the system in question. That was a new approach – rather different from implementing the classical algorithms of, e.g., quantum chemistry, which consume a high amount of computational resources. The first implementation of quantum simulation had been the quantum variant of simulated annealing, a widely used Monte Carlo optimization. In 2011, the Canadian enterprise D-Wave announced to have built the first commercial quantum annealer (Johnson et al., 2011). Others try to implement a universal model of quantum computation using quantum logic gates in superconducting electronic circuits. They tried to reach "quantum supremacy" by means of demonstrating that a programmable quantum device can solve a problem that no classical computer can feasibly solve. In 2019, Google has claimed to have reached this goal (Arute et al., 2019) with its quantum processor for a very special problem, for which the world's largest supercomputer would take thousands of years. New algorithms and software are necessary to exploit the advantages of quantum computing.

Tolcheev (2018) published a bibliometric study on quantum technology including a very broad set of papers (by comprising all papers that use "quantum" in their title, abstract, or keywords). In contrast, this study has a more focused view by including papers from specific technology-relevant subfields. This study does not consider many foundational quantum mechanics/physics applications, quantum chemistry applications, and other quantum related concepts that are too unspecific. These applications and concepts would decrease the precision of the search.

## Methods and Data set

*Data set*

The bibliometric data used in this study are from three sources: (1) the online version of the Web of Science (WoS) database provided by Clarivate Analytics (Philadelphia, Pennsylvania, USA), (2) the bibliometric in-house database of the Max Planck Society (MPG), developed and maintained in cooperation with the Max Planck Digital Library (MPDL, Munich), and (3) the bibliometric in-house database of the Competence Centre for Bibliometrics (CCB, see: http://www.bibliometrie.info/). Both in-house databases were derived from the Science Citation Index Expanded (SCI-E), Social Sciences Citation Index (SSCI), Arts and Humanities Citation Index (AHCI) , Conference Proceedings Citation Index-Science (CPCI-S), and Conference Proceedings Citation Index-Social Science & Humanities (CPCI-SSH) prepared by Clarivate Analytics. The analyses considered publications of the document types “Article”, “Conference Proceeding“, and “Review”. The results are based on 54,598 papers published between 1980 and 2018. The papers had been searched in the online version of the WoS on 25 May 2020 by using 14 subqueries related to QT 2.0. The subqueries have been carefully constructed for sufficient recall and high precision (Bornmann, Haunschild, Scheidsteger, & Ettl, 2019).

*Publication output, citations indicators, and mapping of research topics*

We analyzed the number of papers (full counting) broken down by year, field of QT 2.0, and country. Citation impact analyses are based on time- and field-normalized indicators. We focused on the share of papers belonging to the 10% most frequently cited papers in the corresponding publication year, document type, and subject area. In case of more than one paper with a citation count at the required threshold of 10%, these papers are assigned fractionally to the top 10% publication set. This procedure ensures that there are exactly 10% top 10% papers in each subject area (Waltman & Schreiber, 2013). The top-10% indicator is a standard field-normalized indicator in bibliometrics (Hicks, Wouters, Waltman, de Rijcke, & Rafols, 2015). The citation window relates to the period from publication until the end of 2018.

Besides indicators such as publication and citation counts as measures of scientific activity and impact, techniques of text mining are also used in bibliometric studies. The analysis of keywords in a corpus of publications can identify important research topics and reveal their development over time. This analysis can be managed with the software package VOSviewer (van Eck & Waltman, 2010) which produces networks based on bibliographic coupling. The nodes in these networks are keywords, their size signifies the number of corresponding publications, and the distance between nodes is proportional to their relatedness regarding cited references. Keywords of papers citing similar literature are located closer to each other. The nodes are divided into classes of similarity, displayed by clusters of different colors. The network can be controlled by some adjustable parameters such as minimal cluster size or resolution.

## Results

*Respective share and overall growth of fields*

We retrieved 54.598 publications using the search queries. Table 1 shows the number of papers in the four fields, their percentages of the total number of publications (which add up to more than 100% because of overlaps between the fields), and the percentage of papers belonging to only one field.

**Table 1. Number and percentage of papers in four fields of QT 2.**

| *Field of QT 2.0* | *Number of papers* | *Percentage of the number of distinct papers* | *Number and percentage of one-field-only papers* |
|---|---|---|---|
| i) Q INFO | 16,300 | 29.85% | 9,706 (59.55%) |
| ii) Q METR | 12,531 | 22.95% | 9,766 (77.93%) |
| iii) Q COMM | 13,985 | 25.61% | 9,809 (70.14%) |
| iv) Q COMP | 21,786 | 39.90% | 16,545 (75.94%) |
| Sum of all fields | | 118.77% | 45,826 (83.93%) |

Figure 1 shows the annual publication counts for QT 2.0 and its four fields for the period from 1990 to 2018. The annual numbers of publications on QT 2.0 before 1990 never exceeded a dozen per year. Especially papers about Q METR were published before 1990, which can be explained by the efforts and achievements in manipulation, controlling and measuring of single quantum systems. An exponential growth until about 2000 can be seen in Figure 1, mainly caused by Q METR and Q COMP. From 2000 to 2011, Q COMP is the most strongly represented field, with about twice as many papers as Q INFO and Q COMM.

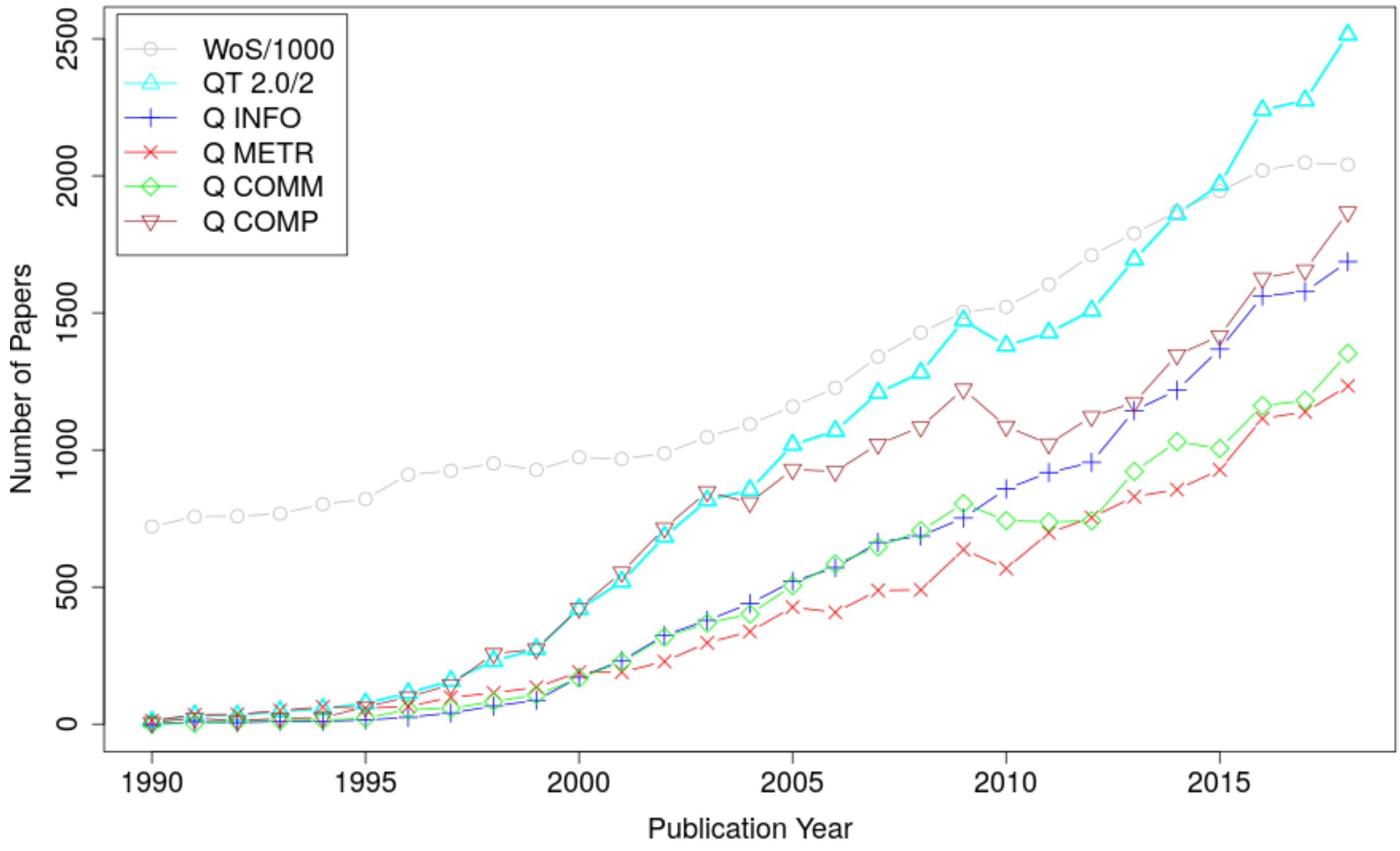

**Figure 1: Annual numbers of publications of QT 2.0 and its four fields between 1990 and 2018 (in earlier years the annual numbers of all fields together never exceed 12) compared to the number of articles, reviews, and conference proceedings in the whole WoS. The numbers for QT 2.0 and the whole WoS are scaled by factors of 2 and 1000, respectively, for better comparison.**

To study the time evolution, we divide the period in three phases: (1) from 1980 to 1999, (2) from 2000 to 2011, and (3) from 2012 to 2018. The numbers of papers in the last two periods are on the same order of magnitude with 24,322 and 28,132. With 2,144 papers, the first pioneering phase has only less than a $10^{th}$ of the paper numbers in the other periods. A measure of the growth of the research fields during the three time periods is the doubling time (see Table 2). The four fields have very similar values as the total QT 2.0. The very short doubling time of two years is characteristic for the first period until 1999, slowing down to four years in the second period and to seven years in the most recent period. The last doubling time is comparable to the 5-6 years which Haunschild, Bornmann, et al. (2016) found for the climate change literature until 2014. However, this time is significantly shorter than the 12-13 years for the overall growth of the WoS records. Bornmann and Mutz (2015) calculated an even higher doubling time of nearly 24 years for the WoS in the time period from 1980 to 2012 by applying a non-linear segmented regression analysis. During the 20 years from 1991 to 2010 the annual number of publications grew by a factor of 42, compared to a factor of ten for the climate change corpus and to a factor of about two for the whole WoS.

**Table 2: Doubling times in years of the three time periods for all QT 2.0 papers and the four fields compared to the whole WoS publication records.**

| *Time period* | *All QT papers* | *Q INFO* | *Q METR* | *Q COMM* | *Q COMP* | *WoS* |
|---|---|---|---|---|---|---|
| 1980-1999 | 2-3 | 1-2 | 3-4 | 1-2 | 1-2 | 7-8 |
| 1980-2011 | 4-5 | 4-5 | 4-5 | 4-5 | 5-6 | 11-12 |
| 1980-2018 | 6-7 | 5-6 | 6-7 | 6-7 | 7-8 | 12-13 |

*Contributing Countries*

Many countries are contributing research on QT 2.0 by collaborating with each other. The 25 top publishing countries (with at least 500 papers published between 2000 and 2018 in QT 2.0) give a similar picture in the four subfields and in QT 2.0 as a whole – with nearly the same countries dominating. The same 22 countries are among the top 25 countries in QT 2.0 and *all four* fields, even when we focus on the 10% most cited papers in QT 2.0 and its fields. For both cases (either all papers or only the top 10% papers), we calculated two numbers: the first number is the difference (%QT - %WoS); positive and negative signs mean more or less publication activity than expected. The second number is the corresponding quotient (%QT / %WoS). The quotient is identical to the so-called Activity Index (AI) which was introduced by Frame (1977). AI is a variant of the Revealed Comparative Advantage (RCA) used in economics (Mittermaier et al., 2017). AIs greater than 1.0 indicate national publication outputs higher than expected (from the whole WoS). Both indicators are presented as radar charts in Figure 2 each containing a plot including all papers and the top 10% papers. In each radar chart, the 22 common countries are denoted by their respective two-letter country codes. The codes start at the top with the country having the most publications in QT 2.0 (US) and descend clockwise. In each radar chart, the dividing values between under and over achievement are marked by a grey dashed line at the value 0 for the difference and 1 for the AI.

The most striking insight from these figure is the very different assessment of the two leading countries with very similar output, the US and China, in comparison with the whole WoS: while the US is less active in QT 2.0 than in other WoS-covered research fields (QT 2.0: Difference = -5.7%, AI = 0.77), China is much more active in QT 2.0 (QT 2.0: Difference = +6.9%, AI = 1.71). The difference is most pronounced in Q COMM (Difference = +15.2%, AI = 2.5). With respect to top 10% papers, the strong research focus of China on QT 2.0 is dampened considerably (QT 2.0: Difference = +1.9%, AI = 1.26; Q COMM: Difference = +7.4%, AI = 2.0).

Figure 2 shows that Austria, Singapore, and Switzerland contribute rather high shares of QT 2.0 research in comparison with their research activities as a whole. Austria has an overall AI of above 2 in QT 2.0 and Q COMP, and of nearly 3 in Q COMM and Q INFO. The AIs even exceeded by focusing on the top 10% papers, leading to values of more than 4. These high AIs are explainable by high activities of research groups in Vienna and Innsbruck concerning quantum teleportation. Singapore has AI values of nearly 3 in Q INFO, Q COMM, and Q COMP. Switzerland's AI value of about 1.6 is mainly caused by a high value of 2.4 in Q COMM.

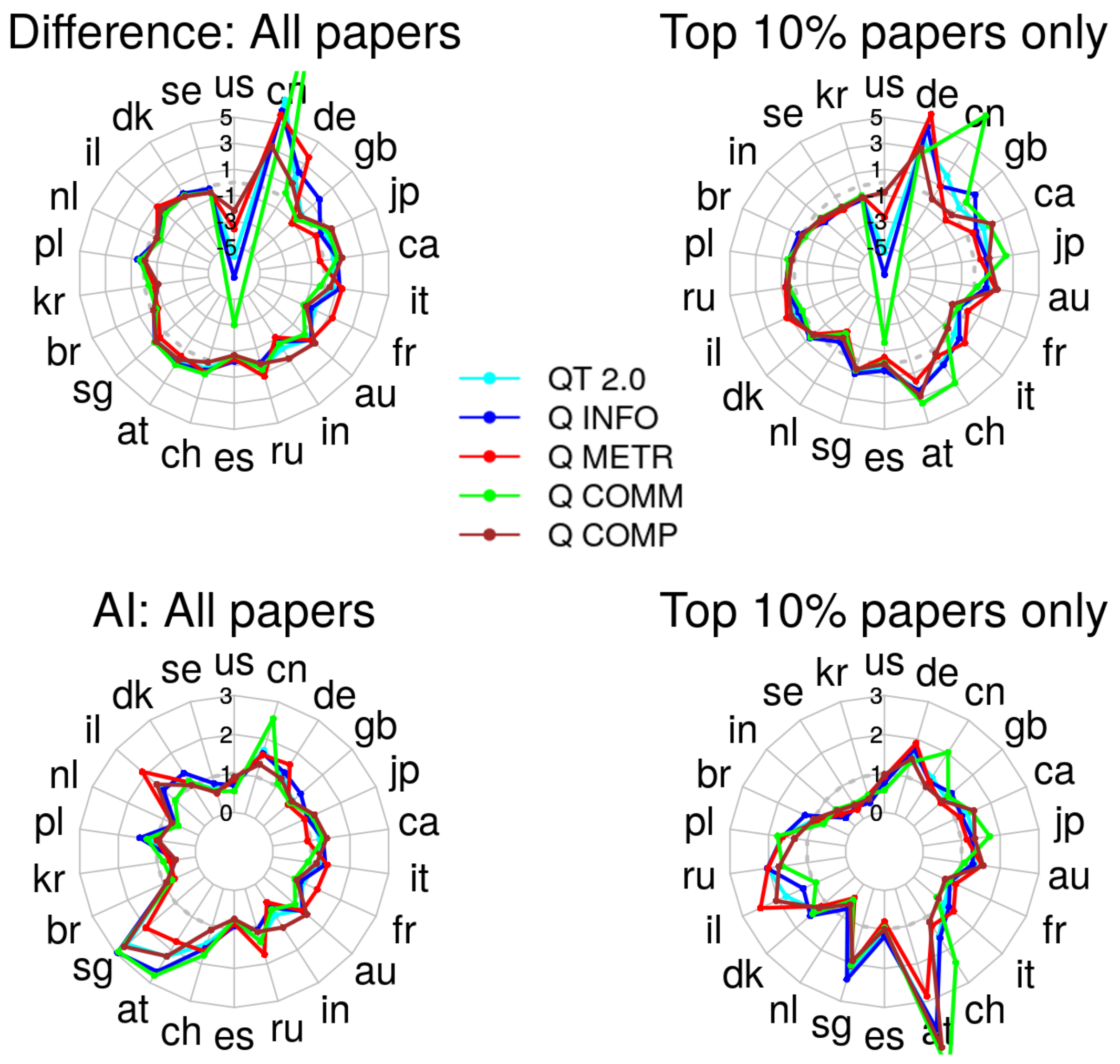

**Figure 2: Radar charts of the differences (upper graphs) and quotients (activity indices, lower graphs) of the national shares of papers in QT 2.0 and its four fields (for the period 2000-2018). On the left side, all papers are included; on the right side, only the top 10% most cited papers in the time period 2000-2016 are included. The 22 countries, which are among the top 25 in all fields and the whole QT 2.0, are denoted by their country codes, and ordered clockwise in descending order of the number of publications. The grey dashed lines at 0 and 1 indicate the expected output of the country. An online version can be viewed at [https://tinyurl.com/yy75v77a](https://tinyurl.com/yy75v77a).**

We investigated to what extent the growth of QT 2.0 research in specific countries was transferred into the area of applications. This can be measured by the share of publications with commercial co-authorships. Since the databases that we used for this study only provide this information for German affiliations, we focused on the subset of publications with authors only from Germany. Because of the very small number of QT 2.0 publications in the first (pioneering) period until 1999, we only consider the second and third period starting from 2000. The share of publications with German commercial co-authorships is between about 4.5 and 5.5% in the whole WoS with German authorship alone from 2000 to 2018. The QT 2.0 share of these publications during the exploration period from 2000 to 2011 is about 2.3%, but during the maturing period from 2012 to 2018 it is nearly twice as high with 4.3%.

*Visualization of research topics and their time evolution*

For the various time periods, we have created keyword maps based on author keywords and keywords plus assigned by Clarivate Analytics. A common thesaurus file (https://tinyurl.com/y35eqaax) was used to unify singular/plural forms of words and synonyms. The minimal number of keyword occurrences is chosen such that about 100 keywords are displayed for each time period. The VOSviewer parameters for the clustering have been chosen to be default values. However, for the minimal cluster size, we used a value of 5 to receive a well interpretable network. All VOSviewer maps are provided as Java-based web-startable versions (Oracle, 2020) via URLs for an interactive inspection by the readers, e.g. by zooming into the clusters.

Figure 3 shows the overall co-occurrence map of 100 keywords occurring at least 298 times for the period from 1980 to 2018. Maps with about 100 keywords usually are a good compromise between keeping readability of the map and displaying most of the content. In the figure, the four fields of QT 2.0 are nicely discernible by the *keywords* in four clusters, whose colors are kept consistent in all networks: (1) Red (Q METR) with the *manipulation* of single *atoms*, *molecules*, and even (*electron*) *spins* as in *quantum dots*, and the *quantum control* using *light fields* of *coherence* (*lasers*) that lead to the *realization* of single *qubits* and of very high precision *quantum clocks*. (2) Brown (Q COMP) with q*uantum computing* and *computers* that build on *quantum circuits* with *logic gates* realized as *trapped ions*, *anyons* or in *NMR* devices. On this hardware, *quantum algorithms* have been implemented that are much in need of *quantum error correction*. (3) Blue (Q INFO) with *quantum entanglement* of *states* that is a major subject of *quantum information science*, also investigating *entropy* and *channel capacity* in a generalization of Shannon's information theory. (4) Green (Q COMM) with q*uantum communication* that is built upon the *quantum teleportation* of *pairs* of *entangled states*, often realized by *single photons*, as a basis for *quantum cryptography* and *quantum key distribution*.

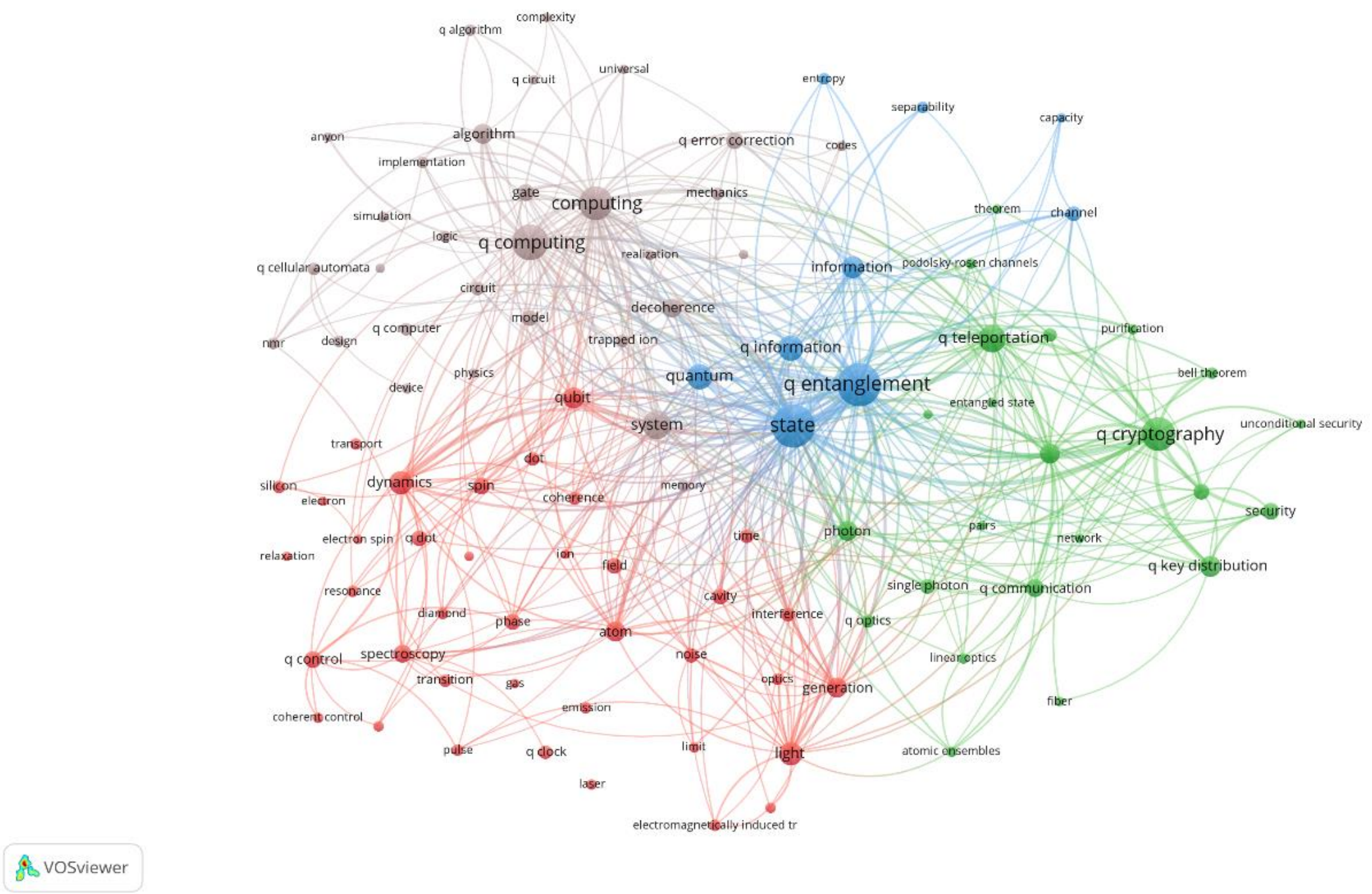

**Figure 3: Co-occurrence map of the top 100 keywords (author keywords and keywords plus) in the period 1980 to 2018 with four topical clusters, using the VOSviewer parameters**

**resolution=1.0 and minimal cluster size=5. (The two biggest unnamed green nodes belong to the keywords communication and key distribution.) For better readability of compound keywords, quantum is abbreviated to q (web-startable version at https://tinyurl.com/y8lvjnyt).**

When we inspect the topic maps for the three partial time periods, we see continuity and persistence of clusters as well as change of focus and occurrence of keywords: from the first to the second time period only 60 out of 99 keywords in the maps are identical. From 1980 to 1999 (https://tinyurl.com/y7c2nxcc) the focus had been on the preparation, manipulation, and control of single quantum systems at the atomic scale and the pioneering work on building materials, devices, and sensors for quantum metrology. From 2000 to 2011 (https://tinyurl.com/ycwqehyu) the focus had, on the one hand, switched to the advanced design of hardware components for real quantum computers and the development of algorithms utilizing quantum properties. On the other hand, the exploitation of quantum effects like entanglement for secure communication using quantum key distribution had become prominent (favorably utilizing quantum optics of single photons). From the second to the third time period, i.e. 2012 to 2018 (https://tinyurl.com/y7c2nxcc), nearly 80% of the keywords remain the same (78 out of 99 and 101 keywords, respectively). There are only slight changes in the main direction of research, but some keywords moved into the new clusters of Q INFO and quantum optics. For example, memory and storage are located in the clusters Q COMP and Q METR in the second period, and are connected with quantum optics in the third period. This connection is probably because of their importance for optical quantum communication networks. The keyword quantum simulation appears only on the third map in the Q COMP cluster. This coincides with the enlarged efforts to build a quantum simulator as the fulfilment of Feynman's vision of a quantum computer (Harris et al., 2018; Johnson et al., 2011).

*Visualization of the geographical distribution of research topics*

Figure 4 shows a combination of the approaches taken in the previous two sections. For the period from 1980 to 2018, we have produced a co-occurrence map of countries (denoted by their two-letter country code with a prefixed “@”) with at least 400 occurrences (multiple co-authorships of the same country on one paper are counted only once) as well as a map of keywords (author keywords and keywords plus assigned by Clarivate Analytics) with at least 300 occurrences. These thresholds lead to the above mentioned top 25 countries and to 104 keywords, sorted into five topical clusters (by using the VOSviewer parameters resolution=1.1 and minimal cluster size=5). Four clusters in the figure can be assigned to the four fields of QT 2.0. About ten countries are assigned to the clusters Q METR and Q INFO, respectively. Three countries are assigned each to Q COMP and Q COMM. In case of Q COMP, India and Iran are mainly connected to the *design* of *logic gates* and *circuits*.

We compare now the assessment of the countries in the radar charts for all QT 2.0 papers in Figure 2 with their placement and connection in Figure 4: the large node of China in the green cluster (Q COMM) mirrors the dominance of China in this field with respect to the total number of papers and its high AI of nearly 3. Germany with the third highest publication output and the highest values in Q METR in the radar charts is located consequently prominently in the red cluster. Germany has significant contributions to quantum optics, and is connected to some other countries in the blue cluster (Q INFO). Q INFO is the field of Germany's second highest AI.

We would like to emphasize two other countries. These countries have with 2.5% a small share of all QT 2.0 papers, but a high AI of about 3 in Q INFO and Q COMM: Singapore has a high AI of about 3 in Q INFO and Q COMM. In the map, consequently, it can be found in the blue cluster of Q INFO connected with *quantum entanglement* and *information* (and with the UK). The country is additionally connected with the green cluster of Q COMM and its major player

China. Austria's activities, especially in Innsbruck and Vienna, are mirrored by its placement in the blue cluster Q INFO. This cluster is strongly connected to *quantum entanglement*. It is also connected to the green Q COMM keywords *communication* and *pairs* of *photons* (quantum optics, orange cluster).

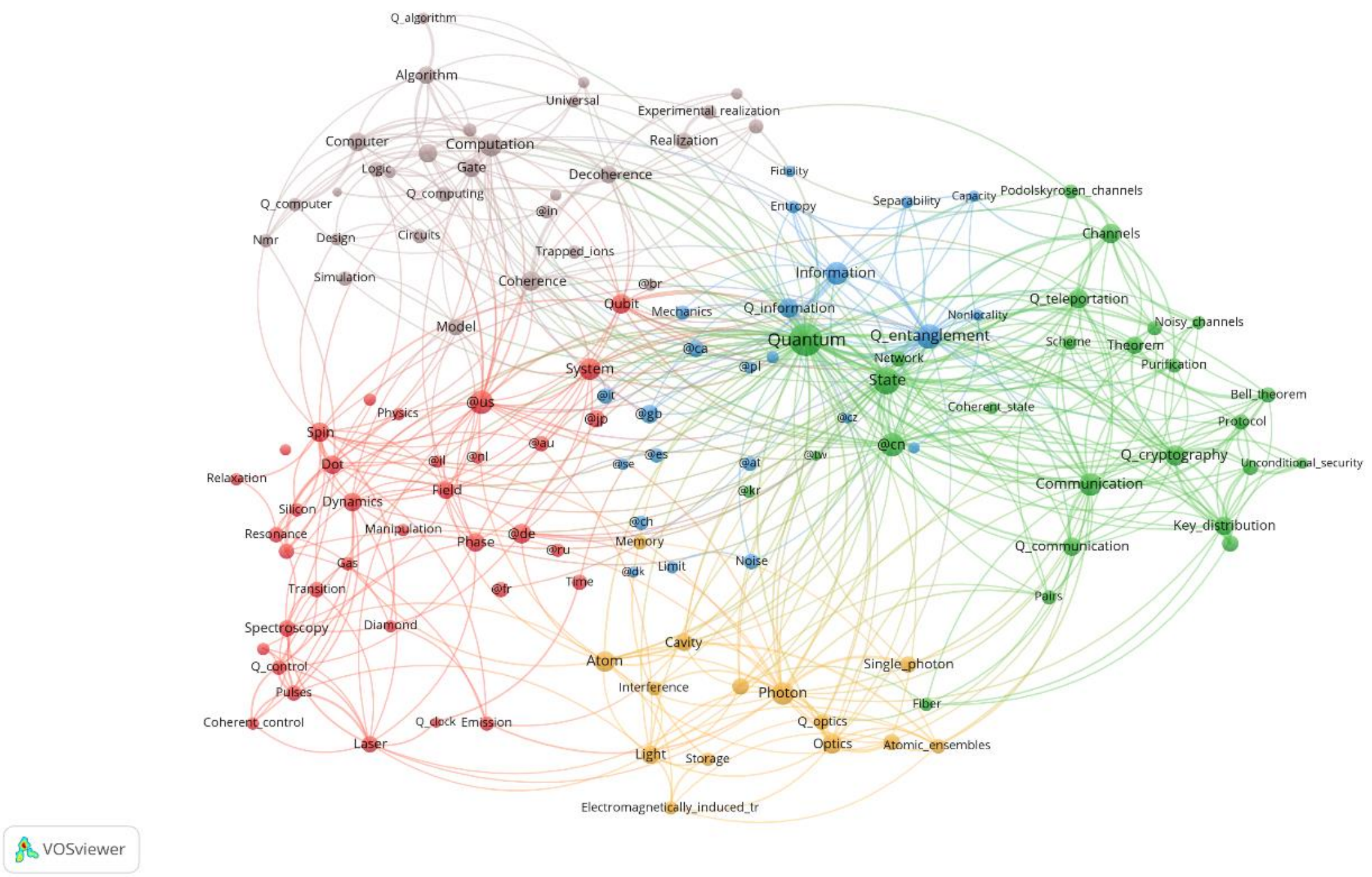

**Figure 4: Co-occurrence map of the top 25 countries (denoted by their two-letter country code with a prefixed "@") with at least 400 occurrences and the top 104 keywords (author keywords and keywords plus) with at least 300 occurrences in the total publication set (from 1980 to 2018) with five topical clusters, using the VOSviewer parameters resolution=1.1 and minimal cluster size=5. For better readability of compound keywords, the term Quantum is abbreviated to Q (web-startable version at [https://tinyurl.com/y978bpzm](https://tinyurl.com/y978bpzm)).**

## Discussion and Conclusions

This bibliometric study on QT 2.0 identified four main subject fields, namely Q INFO, Q METR, Q COMM, and Q COMP. For these four fields, we analyzed their respective share and growth. Of the 54.598 publications in our dataset, the four fields have shares from about one fifth (Q METR) to two fifth (Q COMP) (see Table 1). In the first decade considered here, less than 100 publications have appeared dominated by Q METR with its pioneering works on preparing and controlling single quantum systems. During the second decade (the 1990s) Q COMP joined Q METR in driving the exponential growth, leading to the ongoing dominance of Q COMP in the new millennium (see Figure 1). Between 1980 and 1999, the doubling time of QT 2.0 was between 2 and 3 years; the doubling time of the whole WoS is 7 to 8 years. During the time periods until 2011 and until 2018, respectively, the doubling times were about half as long as in the whole WoS, with 4 to 5 years and 6 to 7 years, respectively (see Table 2). In the most recent decade, therefore, QT 2.0 is a very active research area with a steady exponential development, which is common for mature research fields. We also analyzed the main contributing countries to QT 2.0. We focused on a time period with a substantial annual number of papers from 1990 until 2018. We looked at the top 25 contributing countries in more detail and compared their publication output in QT 2.0 and its four fields to the one expected from the whole WoS (see Figure 2). We visualized the geographical distribution of research topics with a co-occurrence map of countries and keywords in Figure 4. The main result is the sharp contrast of the US and China which are the greatest contributors to QT 2.0. The US shows a much smaller contribution to QT 2.0 than could be expected from their otherwise leading role in science. China has a far over proportional contribution, especially in the field of Q COMM – corroborated by its hub-like function in the topical map. By focussing on highly cited publications, China's share and AI are significantly diminished.
In future studies, the transfer of QT 2.0 research into the area of (commercial) applications might be an interesting research question. In this study, we could only investigate this topic for publications with authors exclusively from Germany.

**Acknowledgments**
The bibliometric data used in this study are from the bibliometric in-house databases of the Max Planck Society (MPG) and the Competence Centre for Bibliometrics (CCB, see: http://www.bibliometrie.info/). The MPG's database is developed and maintained in cooperation with the Max Planck Digital Library (MPDL, Munich); the CCB's database is developed and maintained by the cooperation of various German research organizations. Both databases are derived from the Science Citation Index Expanded (SCI-E), Social Sciences Citation Index (SSCI), Arts and Humanities Citation Index (AHCI) prepared by Clarivate Analytics (Philadelphia, Pennsylvania, USA). We thank one of the reviewers for pointing out the interesting research question concerning commercial applications of QT 2.0.